\documentclass[11pt]{article}

\usepackage{amsmath}
\usepackage{graphicx}
\usepackage{indentfirst}
\usepackage{amssymb}
\usepackage{cite}
\usepackage{color}
\usepackage{subfigure}

\begin{document}

\title{\bf{Thermodynamics with Pressure and Volume under Charged Particle Absorption}}

\date{}
\maketitle

\begin{center}
\author{Bogeun Gwak}$^a$\footnote{rasenis@sejong.ac.kr}\\

\vskip 0.25in
$^{a}$\it{Department of Physics and Astronomy, Sejong University, Seoul 05006, Republic of Korea}\\
\end{center}
\vskip 0.6in

{\abstract
{We investigate the variation of the charged anti-de Sitter black hole under charged particle absorption by considering thermodynamic volume. The variation exactly corresponds to that expected as the first law of thermodynamics. Nevertheless, we find the decrease of the Bekenstein-Hawking entropy for extremal and near-extremal black holes under the absorption to be an irreversible process. This violation of the second law of thermodynamics is only found in the case considering thermodynamic volume. We test the weak cosmic censorship conjecture affected by the violation. Fortunately, the conjecture is still valid. However, extremal and near-extremal black holes do not change their configurations under the absorption. This is quite different from the case without thermodynamic volume.}
}

\thispagestyle{empty}
\newpage
\setcounter{page}{1}

\section{Introduction}\label{sec1}

A black hole has an event horizon through which any particle cannot escape from its gravity, even light. Classically, it implies that no energy or matter can reach an observer located outside of a black hole. However, in consideration of a quantum effect, a small portion of energy can be radiated to outside of the horizon in a black hole spacetime. Then, from the radiated energy, the temperature of the black hole is defined, and the black hole can be treated as a thermal system having this Hawking temperature\cite{Hawking:1974sw,Hawking:1976de}. Further, black holes have an irreducible mass, which is a property that increases in an irreversible process\cite{Christodoulou:1970wf,Bardeen:1970zz,Christodoulou:1972kt,Smarr:1972kt}. The irreducible mass is similar to the entropy in a thermal system, and based on this similarity, the entropy of a black hole is obtained from the irreducible mass. This is the Bekenstein-Hawking entropy of a black hole\cite{Bekenstein:1973ur,Bekenstein:1974ax} and is proportional to the area of the horizon. Using these two thermal properties, temperature and entropy, the laws of thermodynamics are constructed for the black hole as a thermal system.

An interesting conjecture has been applied to the horizon of a black hole. It is called the weak cosmic censorship conjecture in which the horizon of the black hole should cover its inside because a singularity of the black hole geometry is located at the center of the spacetime\cite{Penrose:1964wq,Penrose:1969pc}. This is an inevitable conjecture to save the causality of the spacetime from a naked singularity. Although the conjecture is generally suggested for black holes, its validity should be tested for each case because there is no general procedure to prove it. Moreover, the validity of the conjecture depends on the methods of investigation. For the Kerr black hole, the conjecture is valid under adding a particle\cite{Wald:1974ge}. However, the near-extremal Kerr black hole can be overspun beyond its extremal bound by a particle, so the conjecture is invalid\cite{Jacobson:2009kt}. To resolve this invalidity, in consideration of the self-force effect, the conjecture is known to be valid for the Kerr black hole\cite{Barausse:2010ka,Barausse:2011vx,Colleoni:2015ena}. The Reissner-Nordstr\"{o}m (RN) black hole was tested for validity of the conjecture with back-reaction effect\cite{Hubeny:1998ga,Isoyama:2011ea}. Furthermore, various investigations have been conducted on the conjecture for not only black holes in Einstein's gravity, but also anti-de Sitter (AdS), lower-dimensional, or higher-dimensional black holes\cite{Colleoni:2015afa,BouhmadiLopez:2010vc,Rocha:2011wp,Gao:2012ca,Rocha:2014gza,Rocha:2014jma,McInnes:2015vga,Hod:2016hqx,Natario:2016bay,Horowitz:2016ezu,Duztas:2016xfg,Gwak:2017icn,Sorce:2017dst}. From a thermodynamic point of view, the conjecture is quite consistently related to the laws of thermodynamics. If the entropy of the black hole increases as ensured by the second law for an irreversible process, the horizon can cover the inside of a black hole as the conjecture suggests. In addition, in the process, the variation of a black hole is consistent with the first law of thermodynamics under a particle absorption\cite{Gwak:2015fsa,Gwak:2016gwj}.

The thermal properties have an important role in AdS spacetime. The gravity theory in $D$-dimensional AdS spacetime is associated with the conformal field theory (CFT) defined on the boundary of the AdS spacetime. This is the AdS/CFT correspondence\cite{Maldacena:1997re,Gubser:1998bc,Witten:1998qj,Aharony:1999ti}. Under this duality, the thermal properties of the AdS black hole are also found in the dual CFT, so that the CFT is given at finite temperature\cite{Witten:1998zw}. Currently, there are various applications of AdS/CFT duality. One of the representative applications is anti-de Sitter/quantum chromodynamics (AdS/QCD) correspondence\cite{Babington:2003vm,Kruczenski:2003uq,Sakai:2005yt,Erlich:2005qh}. Another is anti-de Sitter/condensed matter theory (AdS/CMT) correspondence\cite{Hartnoll:2008vx,Hartnoll:2008kx}. Because each solution of black holes is based on various gravity theories, its dual theory and physical interpretation depend on the black hole concerned. Further, the instability of black holes in perturbation or thermodynamics affects the states of the dual CFT. For example, the charged AdS black hole is mainly related to the AdS/CMT applications. In the (2+1)-dimensional charged AdS black hole, its dual theory is associated with the holographic superconductor\cite{Jensen:2010em,Andrade:2011sx,Chang:2014jna}. In addition, Fermi-Luttinger liquids is a model having resemblance to its dual theory\cite{Hung:2009qk,Davison:2013uha}.

The cosmological constant is a parameter which plays an important role in determining the asymptotic topology of a black hole spacetime. In the action of Einstein gravity, the cosmological constant is fixed, so it is set to a constant value at any time. Recently, various interesting studies have been conducted on the thermodynamics of black holes wherein the cosmological constant has been set to a dynamic variable and interpreted as a pressure. In fact, the cosmological constant as a dynamic variable was first considered a long time ago\cite{Teitelboim:1985dp,Brown:1988kg}. Furthermore, the pressure of the black hole spacetime is associated with the cosmological constant\cite{Caldarelli:1999xj, Padmanabhan:2002sha}, and its thermal conjugate is found to be a thermodynamic volume\cite{Dolan:2010ha,Cvetic:2010jb}. Under the cosmological constant as a dynamic variable, the mass of the black hole corresponds to the enthalpy of the black hole system\cite{Kastor:2009wy}. Owing to the pressure-volume contribution, the first law of thermodynamics is extended to have $PV$ term\cite{Dolan:2011xt}. This considerably affects the thermal phase of the black hole; various phenomena have been already investigated such as Van der Waals fluids, reentrant phase transitions, and holographic heat engines\cite{Kubiznak:2012wp,Johnson:2014yja,Karch:2015rpa,Kubiznak:2016qmn,Hennigar:2017apu,Wei:2017vqs}.

In this work, we prove that the variation of the $D$-dimensional charged AdS black hole including four dimensions follows the first law of thermodynamics considering the thermodynamic volume term via the charged particle absorption. Further, we investigate the second law of thermodynamics. Because particle absorption is an irreversible process, the entropy of the black hole should increase. It has been already proved that the satisfaction of the first law of thermodynamics is a necessary condition to ensure the second law of thermodynamics under particle absorption\cite{Gwak:2015fsa,Gwak:2016gwj} in the case of the non-thermodynamic volume term. Nevertheless, if the second law of thermodynamics is not valid under the absorption, it would be the first violation of the second law of thermodynamics under the correct first law of thermodynamics, which would only be seen in the case considering the pressure and volume term. The second law of thermodynamics plays an important role in physical processes such as the collision of black holes\cite{Gwak:2017zwm}. Considering the thermodynamic volume term, we assume the cosmological constant as a dynamic variable in the metric of the black hole. Under this assumption, we test the weak cosmic censorship conjecture by the charged particle absorption. Moreover, in the Einstein-Maxwell action, the cosmological constant is not a dynamic variable, so we cannot test the conjecture under the level of equations of motion. Thus, the particle absorption is almost the only method to investigate the conjecture from the variation of the black hole including the pressure and volume term.

This paper is organized as follows. In section~\ref{sec2}, the charged AdS black hole is introduced, and the laws of thermodynamics are presented along with the dynamic cosmological constant. In section~\ref{sec3}, we establish the first law of thermodynamics under the charged particle absorption. Further, the second law of thermodynamics is shown to be violated in specific cases. In section~\ref{sec4}, we describe the investigation of the weak cosmic censorship conjecture for the extremal and near-extremal black holes. In section~\ref{sec5}, we briefly summarize our results.

\section{Thermodynamic Volume in Charged AdS Black Hole}\label{sec2}

The Einstein-Maxwell action with the cosmological constant in the $D$-dimensional spacetime is given as
\begin{align}\label{eq:EMCaction}
S=-\frac{1}{16\pi}\int d^D x \sqrt{-g} \left(R-F_{\mu\nu}F^{\mu\nu}-2\Lambda\right).
\end{align}
where the spacetime dimensions are denoted as $D$ and include four dimensions. Maxwell field strength $F_{\mu\nu}$ and electric potential $A_\mu$ are$A_\mu$ are
\begin{align}
F_{\mu\nu}=\partial_\mu A_\nu - \partial_\nu A_{\mu}, \quad  A=-\frac{Q}{r^{D-3}}dt.
\end{align} 
The equations of motion from Eq.\,(\ref{eq:EMCaction}) have a static solution for the charged AdS black hole. The metric of the black hole is in $D$-dimensional spacetime
\begin{align}\label{eq:metric01}
ds^2 = - f(r)dt^2 + f(r)^{-1}dr^2 + r^2 d\Omega_{D-2},\quad f(r) = 1-\frac{2M}{r^{D-3}}+\frac{Q^2}{r^{2D-6}}+\frac{r^2}{\ell^2},
\end{align}
where the $D-2$-dimensional sphere is analytically denoted as 
\begin{align}
d\Omega_{D-2}=\sum^{D-2}_{i=1}\left( \prod_{j=1}^{i} \sin^2\theta_{j-1}\right)d\theta_i^2,\quad \theta_0\equiv\frac{\pi}{2},\quad \theta_{D-2}\equiv\phi.
\end{align}
The metric components are determined in terms of mass and charge parameters $M$ and $Q$ with an AdS radius $\ell$ in Eq.\,(\ref{eq:metric01}). These parameters are proportional to the mass $M_\text{b}$, electric charge $Q_\text{b}$, and cosmological constant $\Lambda$\cite{Tian:2010gn}.
\begin{align}
M_\text{b}=\frac{(D-2)\Omega_{D-2}}{8\pi}M,\quad Q_\text{b}=\frac{(D-2)\Omega_{D-2}}{8\pi}Q,\quad \Lambda=-\frac{(D-1)(D-2)}{2\ell^2},
\end{align}
where we set $G=1$ and $\hbar=1$ in all dimensions for simplicity. The thermal properties can be defined on the horizon $r_\text{h}$ of the black hole. Hawking temperature, Bekenstein-Hawking entropy, and electric potential are given as
\begin{align}\label{eq:temperature02}
T_\text{h}=\frac{1}{2\pi \ell^2 }\left(r_\text{h}-\frac{(D-3)Q^2\ell^2}{r_\text{h}^{2D-5}}+\frac{(D-3)M\ell^2}{r_\text{h}^{D-2}}\right),\quad S_\text{h}=\frac{A_\text{h}}{4}=\frac{\Omega_{D-2}r_\text{h}^{D-2}}{4},\quad \Phi_\text{h}=\frac{Q}{r_\text{h}^{D-3}}.
\end{align}
Then, the thermodynamic laws can be constructed for the black hole. Recently, an interesting approach has been followed to treat the cosmological constant as a thermodynamic variable. From this point of view, the cosmological constant is not a fixed value. Its actual value can be obtained from the vacuum expectation value of the theory considered, so that it can vary under a perturbation\cite{Cvetic:2010jb}. Although the variation of the cosmological constant is not concrete in the Lagrangian theory, the cosmological constant as a thermodynamic variable represents quite consistent behaviors with other thermodynamic variables\cite{Dolan:2010ha,Dolan:2011xt}. In this extended thermodynamics, the cosmological constant plays the role of pressure $P$, and its conjugate variable is thermodynamic volume of the black hole $V_\text{b}$. The definitions of thermodynamic pressure and volume are in $D$-dimensional AdS spacetime\cite{Dolan:2013ft}
\begin{align}\label{eq:volume03}
P=-\frac{\Lambda}{8\pi}=\frac{(D-1)(D-2)}{16\pi \ell^2},\quad V_\text{b}=\frac{\Omega_{D-2}}{D-1}r_\text{h}^{D-1}.
\end{align}
When we consider the pressure term in the laws of thermodynamics, the key difference is that the mass is now enthalpy in the first law of thermodynamics\cite{Kastor:2009wy,Cvetic:2010jb}. Thus, the first law of thermodynamics determines the infinitesimal change of the mass of the black hole as\cite{Dolan:2012jh,Dolan:2013ft}
\begin{align}
dM_\text{b}=T_\text{h}dS_\text{h}+\Phi_\text{h}dQ+V_\text{h}dP,
\end{align}
where the $M_\text{b}$ plays a role as an enthalpy. The enthalpy is related to the internal energy of the black hole $U_\text{b}$ and the $PV_\text{b}$ term as
\begin{align}
M_\text{b}=U_\text{b}+PV_\text{b}.
\end{align}
Therefore, the variation of the mass causes rebalancing not only of the horizon and electric charge, but also the AdS radius in $PV_\text{b}$ term. In the following section, we will investigate the change in the black hole by the charged particle absorption when the AdS radius is assumed to be infinitesimally varied because of it.

\section{Thermodynamics with Pressure and Volume under Charged Particle Absorption}\label{sec3}

We assume that the charged AdS black hole is varied by absorbing a charged particle. When the black hole absorbs the particle, the conserved quantities of the particle can perturb both the mass and charge of the black hole, and the AdS radius is subordinated to these changes owing to the contribution of thermodynamic pressure and volume. To analyze this charged particle absorption, we will obtain the relation between conserved quantities of the particle, because the conserved quantities of the black hole are changed as much as those of the particle. Then, Hamiltonian of the charged particle under an electric potential $A_\mu$ is given as
\begin{align}
\mathcal{H}=\frac{1}{2}g^{\mu\nu}(p_\mu-q A_\mu)(p_\nu-q A_\nu),
\end{align}
of which Hamiltonian equations are separable under Hamilton-Jacobi action\cite{Carter}. The momentum $p_\mu$ is obtained in terms of a partial derivative of Hamilton-Jacobi action that
\begin{align}\label{eq:hjaction}
S=\frac{1}{2}m^2 \lambda -Et+L\phi+S_r(r)+\sum_{i=1}^{D-3}S_{\theta_i}(\theta_i),\quad p_\mu=\partial_\mu S.
\end{align}
The Hamilton-Jacobi action describes a particle having a mass $m^2$, and the affine parameter is $\lambda$. The conserved quantities $E$ and $L$ are assumed from the translation symmetries of the metric in Eq.\,(\ref{eq:metric01}). Owing to $D$-dimensional solution, the black hole includes a $D-2$-dimensional sphere $\Omega_{D-2}$. The angular momentum $L$ is defined as the conserved quantity from the translation symmetry of the last angle coordinate of $\Omega_{D-2}$. Thus, the summation in Eq.\,(\ref{eq:hjaction}) runs from $i=1$ to $D-3$. To solve Hamilton-Jacobi equations, we can use the inverse metric including $D-2$-dimensional sphere, so
\begin{align}
g^{\mu\nu}\partial_\mu\partial_\nu = -f(r)^{-1}(\partial_t)^2+f(r)(\partial_r)^2+r^{-2}\sum^{D-2}_{i=1}\left( \prod_{j=1}^{i} \sin^{-2}\theta_{j-1}\right)(\partial_{\theta_i})^2.
\end{align}
The Hamilton-Jacobi equation is
\begin{align}
-2\frac{\partial S}{\partial \lambda}=-m^2=&-f(r)^{-1}(-E-qA_t)^2+f(r)(\partial_r S_r(r))^2\\
&+r^{-2}\sum^{D-3}_{i=1}\left( \prod_{j=1}^{i} \sin^{-2}\theta_{j-1}\right)(\partial_{\theta_i}S_{\theta_i}(\theta_i))^2+r^{-2}\left(\prod_{j=1}^{D-2} \sin^{-2}\theta_{j-1}\right)(L)^2,\nonumber
\end{align}
which is divided by separate variables, $\mathcal{K}$ and $R_i$.
\begin{align}
\mathcal{K}=-m^2 r^2+\frac{r^2}{f(r)}(-E+\frac{Qq}{r^{D-3}})^2-r^2 f(r)(\partial_r S_r(r))^2,\quad R_i^2=(\partial_i S_{\theta_i}(\theta_i))^2 + \sin^2\theta_i R_{i+1}^2,
\end{align}
where two variables are defined that
\begin{align}
\mathcal{K}=R_1^2,\quad L=R_{D-2}.
\end{align}
Then, we can determine the entire equations of motion. The radial- and $\theta$-directional equations are sufficient to obtain the relation between energy and electric charge of the particle. The momenta of the particle are
\begin{align}\label{eq:eoms01}
p^r\equiv \frac{\partial r}{\partial \lambda}= f(r) \sqrt{-\frac{\mathcal{K}+m^2r^2}{r^2 f(r)}+\frac{1}{f(r)^2}\left(E-\frac{Qq}{r^{D-3}}\right)^2},\quad p^\theta\equiv \frac{\partial \theta}{\partial \lambda}=\frac{1}{r^2}\sqrt{\mathcal{K}-\sin^2\theta_1 R_2^2}.
\end{align}
We attempt to determine the variation of the black hole which absorbs a charged particle. The particle is supposed to be absorbed to the black hole when it passes through the outer horizon $r_\text{h}$, because the conserved quantities of the particle are not distinguishable anymore from those of the black hole at that moment by an observer outside of the horizon. By removing the separate variable $\mathcal{K}$ in Eq.\,(\ref{eq:eoms01}), we obtain the relation between conserved quantities and momenta for a given radial location $r$. Then, at the outer horizon $r_h$, conserved quantities of the particle are absorbed to those of the black hole. At the limit of the outer horizon, the energy relation between conserved quantities and momenta is obtained as
\begin{align}\label{eq:dispersion01}
E=\frac{Q}{r_\text{h}^{D-3}}q+|p^r|,
\end{align}
in which various dependencies between variables are reduced to this simple relation. A positive sign is required in front of the $|p^r|$ term. In the positive flow of time, the particle comes into the black hole. At this moment, the energy of the particle should be defined as a  positive value, so that the signs in front of $E$ and $|p^r|$ are the same and positive\cite{Christodoulou:1970wf,Christodoulou:1972kt}. Note that we consider the energy dependent on the electric potential term. However, the potential is independent of the flow of time and only related to the interaction between particle and black hole. Thus, the total value of energy under the sum of the potential is not important, and we simply choose the positive sign in front of $|p^r|$. 

Absorbing the charged particle, the black hole is varied by the same quantity as that of the particle, assuming no loss of conserved quantities in this process. This is supported by the change of the black hole following the first law of thermodynamics. The charge of the particle $q$ is coincident to the change of the charge of the black hole $dQ_\text{b}$. The energy of the particle is only given as $q$ and $|p^r|$ at the horizon in Eq.\,(\ref{eq:dispersion01}), so we must find a thermodynamic potential of which the variables also change by $q$ and $|p^r|$. If we assume that the energy of the particle changes the internal energy of the black hole, the internal energy is given as $U_\text{b}(Q_\text{b},S_\text{b},V_\text{b})$, so that its variation will be given as $dQ_\text{b}$, $dS_\text{b}$, and $dV_\text{b}$. Fortunately, these variables will be denoted in terms of those of the particle. The energy and electric charge of the particle become
\begin{align}
E=dU_\text{b}=d(M_b-PV_\text{b}),\quad q=dQ_\text{b}.
\end{align}
Then, the energy relation in Eq.\,(\ref{eq:dispersion01}) is rewritten as
\begin{align}\label{eq:dispersion02}
dU_\text{b}=\frac{Q}{r_\text{h}^{D-3}}dQ_b+|p^r|.
\end{align}
Thus, the charged particle changes the black hole as much as $(dU_\text{b},dQ_\text{b})$, and the change of the black hole volume induces the change of its conjugate variable, pressure. Under the charged particle absorption, the changed variables are $(dM_\text{b},dQ_\text{b}, d\ell)$. The other variables depend on these. To rewrite Eq.\,(\ref{eq:dispersion02}) to the first law of thermodynamics, we need to find $dS_\text{h}$ changed by the absorption. Under the variation,
\begin{align}
dS_\text{h}=\frac{1}{4}(D-2)\Omega_{D-2}r_\text{h}^{D-3}dr_\text{h},
\end{align}
where the change of the outer horizon $dr_\text{h}$ should be rewritten as independent variables such as $(q,|p^r|)$ of the particle. The particle absorption varies the function $f(r)$, and its change is the reason of moved outer horizon $r_\text{h}+dr_\text{h}$. The infinitesimally moved outer horizon $dr_\text{h}$ satisfies
\begin{align}\label{eq:function05}
df_\text{h}=\frac{\partial f_\text{h}}{\partial M_\text{b}}dM_\text{b}+\frac{\partial f_\text{h}}{\partial Q_\text{b}}dQ_\text{b}+\frac{\partial f_\text{h}}{\partial \ell}d\ell+\frac{\partial f_\text{h}}{\partial r_\text{h}}dr_\text{h}=0,\quad f_\text{h}=f(M_\text{b},Q_\text{b},\ell,r_\text{h}),
\end{align} 
with
\begin{align}
\frac{\partial f_\text{h}}{\partial M_\text{b}}&=-\frac{16\pi}{(D-2)\Omega_{D-2}r_\text{h}^{D-3}},\quad \frac{\partial f_\text{h}}{\partial Q_\text{b}}=\frac{16\pi Q}{(D-2)\Omega_{D-2}r_\text{h}^{2D-6}},\\
\frac{\partial f_\text{h}}{\partial \ell}&=-\frac{2 r_\text{h}^2}{\ell^3},\quad \frac{\partial f_\text{h}}{\partial r_\text{h}}=-\frac{(2D-6)Q^2}{r_\text{h}^{2D-5}}+\frac{2(D-3)M}{r_\text{h}^{D-2}}+\frac{2r_\text{h}}{\ell^2}.
\nonumber
\end{align}
In addition, the energy relation in Eq.\,(\ref{eq:dispersion02}) is rewritten in terms of the enthalpy
\begin{align}\label{eq:dispersion03}
dM_\text{b}-d(PV_\text{h})=\frac{8\pi Q_\text{b}}{(D-2)\Omega_{D-2}r^{D-3}}dQ_\text{b}+|p^r|.
\end{align}
By combining Eqs.\,(\ref{eq:function05}) and (\ref{eq:dispersion03}), we can remove the $d\ell$ term. Then, interestingly, except for the $|p^r|$ and $dr_\text{h}$ variables, the others are removed. The change of the outer horizon becomes
 \begin{align}\label{eq:variables02}
dr_\text{h}=\frac{16 \pi r_\text{h}^{4}\ell^2 |p^r|}{\Omega_{D-2}(D-2)(D-3)(r_\text{h}^{D+2}-2 M r_\text{h}^3\ell^2 +2r_\text{h}^{D}\ell^2)}.
\end{align}
Therefore, under the energy relation, the variations of entropy, and thermodynamic volume of the black hole is obtained as
\begin{align}\label{eq:variables01}
dS_\text{h}&=\frac{4 \pi r_\text{h}^{D+1}\ell^2 |p^r|}{(D-3)(r_\text{h}^{D+2}-2 M r_\text{h}^3\ell^2 +2r_\text{h}^{D}\ell^2)},\\
dV_\text{h}&=\frac{16 \pi r_\text{h}^{D+1}\ell^2 |p^r|}{(D-2)(D-3)(r_\text{h}^{D+2}-2 M r_\text{h}^3\ell^2 +2r_\text{h}^{D}\ell^2)}.\nonumber
\end{align}
Incorporating Eqs.\,(\ref{eq:temperature02}), (\ref{eq:volume03}), and (\ref{eq:variables01}), the energy relation of Eq.\,(\ref{eq:dispersion02}) becomes the first law of thermodynamics that
\begin{align}
dU_\text{h}=\Phi_\text{h}dQ_b + T_\text{h} dS_\text{h}-PdV_\text{h}.
\end{align}
Because the mass of the black hole is already defined to the enthalpy of the black hole, the internal energy can be exchanged with the enthalpy by the Legendre transformation, so
\begin{align}
dM_\text{b}=T_\text{h}dS_\text{h}+\Phi_\text{h}dQ+V_\text{h}dP.
\end{align}
Thus, we prove the coincidence between the variation of the $D$-dimensional charged black hole and the first law of thermodynamics under the charged particle absorption.

The second law of thermodynamics expects the increase of the entropy of the black hole in an irreversible process. As the charged particle absorption is an irreversible process, the entropy becomes greater than before the absorption. The validity of this statement is easily checked by the sign of $dS_\text{h}$ in Eq.\,(\ref{eq:variables01}). Moreover, we obtain the violation of the second law of thermodynamics in parameter ranges. Specifically, the denominator of $dS_\text{h}$ has a negative value for the extremal black hole, 
\begin{align}\label{eq:extremaldS01}
r_\text{h}^{D+2}-2 M r_\text{h}^3\ell^2 +2r_\text{h}^{D}\ell^2=-\frac{(D-1)r_\text{h}^{D+2}}{(D-3)}<0,
\end{align}
which means that the entropy of the black hole decreases at least for the extremal case for all dimensions, $D\geq 4$.
\begin{figure}[h] \centering\subfigure[{$Q-M$ diagram for $D=4$.}] {\includegraphics[scale=0.49,keepaspectratio]{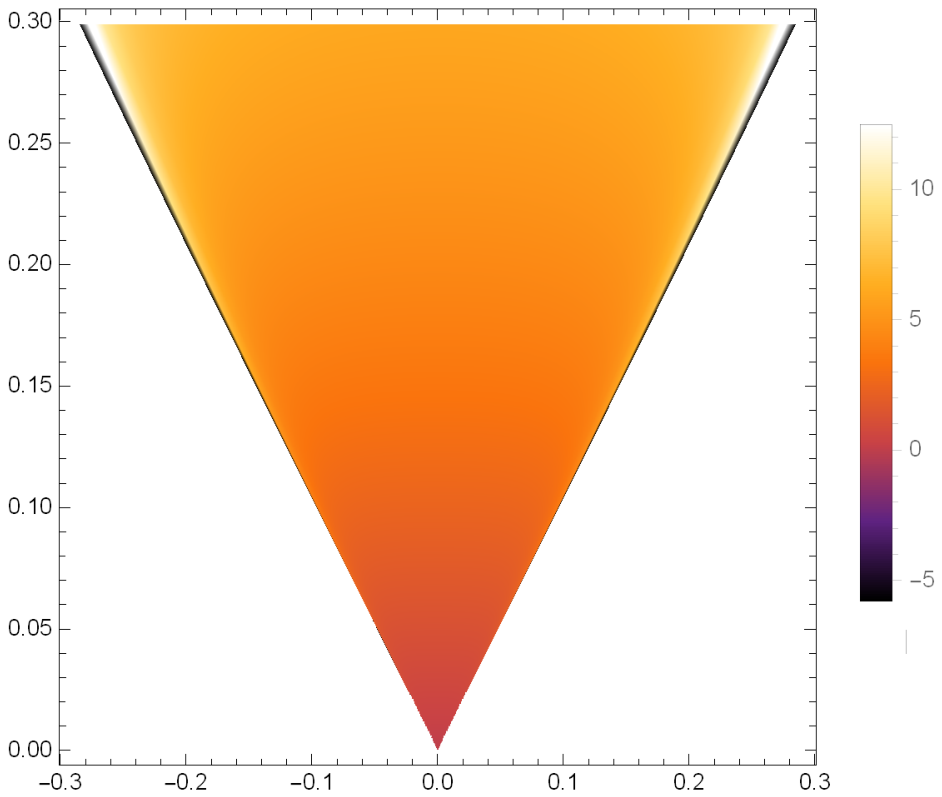}} \quad \centering\subfigure[{$Q-M$ diagram for $D=5$.}] {\includegraphics[scale=0.49,keepaspectratio]{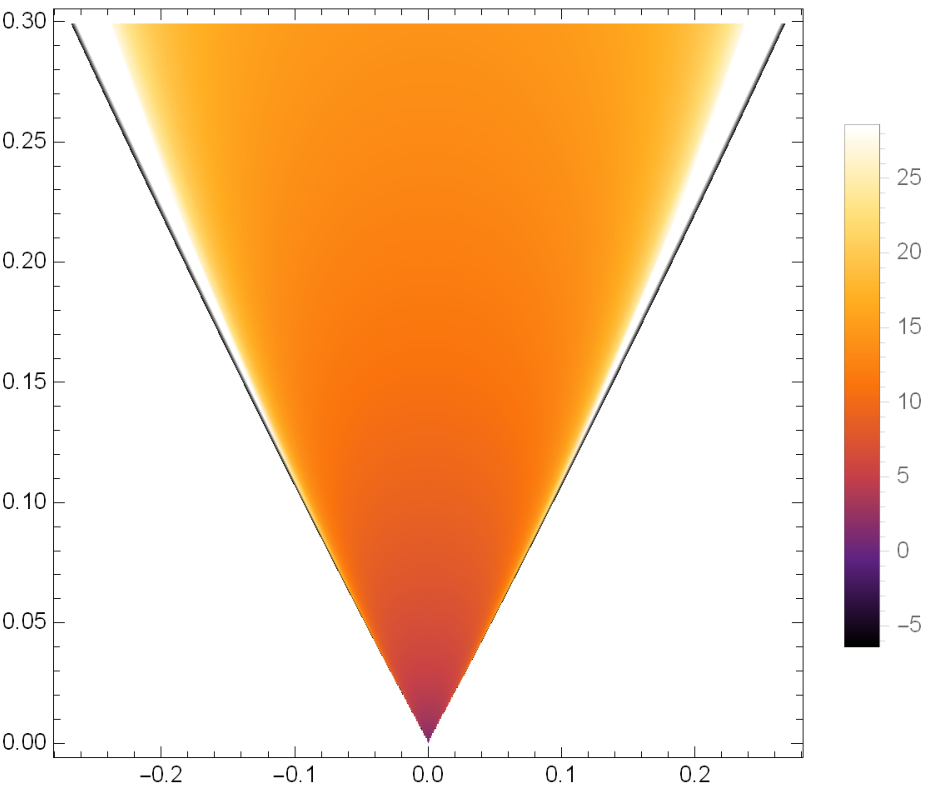}} \quad \centering\subfigure[{$Q-M$ diagram for $D=6$.}] {\includegraphics[scale=0.49,keepaspectratio]{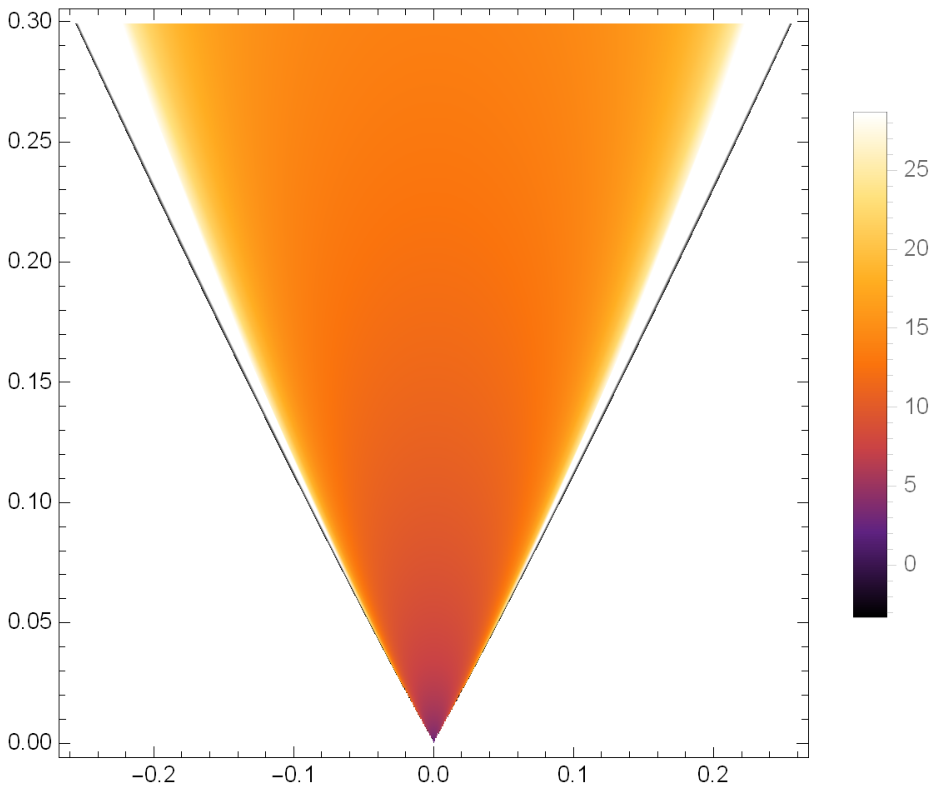}}\\
\caption{{\small The scaled $dS_\text{h}$ in $Q-M$ diagrams with $\ell=1$.}}
\label{fig:f1a}
\end{figure}
Therefore, the second law of thermodynamics is violated under the consideration of $PV_\text{b}$ term for the charged particle absorption. This behavior is interesting and observable only with the $PV_\text{b}$ term in the first law of thermodynamics. Note that without the $PV$ term, the second law of thermodynamics is always ensured under satisfying the first law of thermodynamics\cite{Gwak:2015fsa}. Using $dS_\text{h}$ in Eq.\,(\ref{eq:variables01}), we investigate parameter ranges within which this violation occurs. The area of the violation depends on the spacetime dimensions. The parameter space $(Q,M)$ is limited due to the extremal condition for $D$-dimensional black holes in Fig.\,\ref{fig:f1a} in which negative maxima is scaled to have around $-5$.   As already shown in Eq.\,(\ref{eq:extremaldS01}), the decrease of the entropy appears in ranges close to the extremal black holes. Then, as the black hole electrically neutralizes more than the extremal ones, the entropy increases. The area of the decrease is obtained in higher dimensions as shown in Fig.\,\ref{fig:f1a} from (a) to (c). In Fig.\,\ref{fig:f1a} (a), (b), and (c), we show parameter spaces of higher-dimensional black holes from four to six dimensions. The range of the violation appears in extremal and near-extremal condition. This also implies that the entropy needs a correction term to resolve the violation when we consider thermodynamic pressure and volume.

\section{Weak Cosmic Censorship Conjecture with Pressure and Volume}\label{sec4}

In consideration of thermodynamic volume, the charged particle absorption can reproduce the first law of thermodynamics in terms of enthalpy. However, although the particle absorption is an irreversible process, we find the violation of the second law of thermodynamics in the process, and the entropy of the black hole decreases for extremal and near-extremal black holes. Without $PV_\text{b}$ term, the second law of thermodynamics is satisfied to validate the weak cosmic censorship conjecture under the absorption. Thus, owing to the violation of the second law with the $PV_\text{b}$ term, we can expect that the cosmic censorship is affected by the term\cite{Gwak:2016gwj}. 

\begin{figure}[h] \centering\subfigure[{$(Q,M,\ell)$ surface satisfying $\delta=0$.}] {\includegraphics[scale=0.5,keepaspectratio]{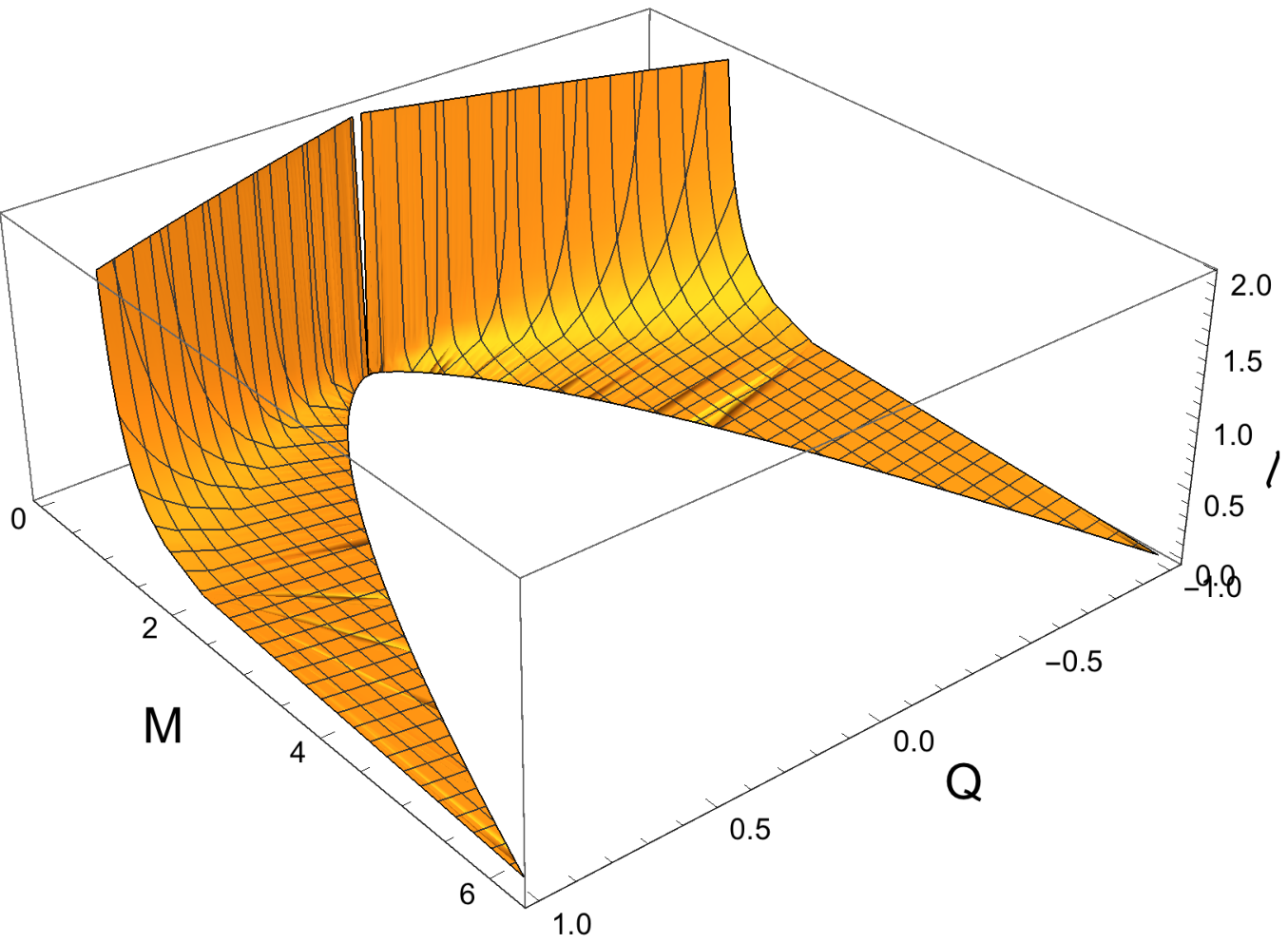}} \quad \centering\subfigure[{$Q-M$ lines in the $\ell=1$ slice for values of $\delta$.}] {\includegraphics[scale=0.8,keepaspectratio]{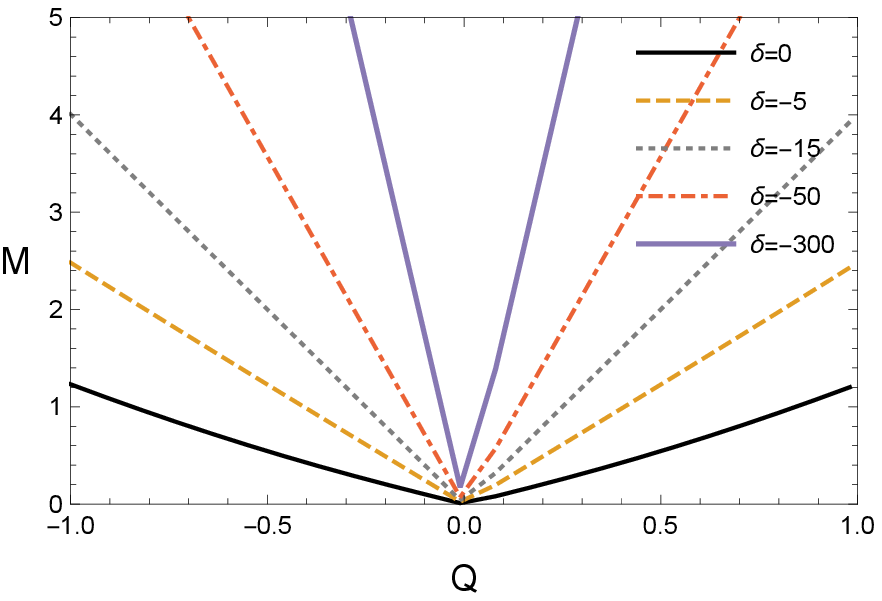}}\caption{{\small $(Q,M,\ell)$ surface and $Q-M$ diagram}} \label{fig:f2b}\end{figure}

As the violation of the second law of thermodynamics occurs in extremal and near-extremal black holes, the changes of extremal and near-extremal black holes become very different from those of non-extremal ones. This change can be estimated from a behavior of the function $f(r)\equiv f(M_\text{b},Q_\text{b},\ell,r)$ in the metric of Eq.\,(\ref{eq:metric01}). The function $f(M_\text{b},Q_\text{b},\ell,r)$ of the black hole has only one minimum point $r_\text{min}$ which satisfies
\begin{align}\label{eq:cosmicExtremal}
f(M_\text{b},Q_\text{b},\ell,r)|_{r=r_\text{min}}\equiv f_\text{min}=\delta\leq 0,\quad \partial_{r} f(M_\text{b},Q_\text{b},\ell,r)|_{r=r_\text{min}}\equiv f'_\text{min}=0,
\end{align}
with 
\begin{align}
 (\partial_{r})^2 f(M_\text{b},Q_\text{b},\ell,r)|_{r=r_\text{min}}>0.
\end{align}
The minimum value of the function $f(M_\text{b},Q_\text{b},\ell,r)$ is $\delta$, and $\delta=0$ for the extremal black hole. In addition, the inner and outer horizons are located around the minimum point. We explicitly denote variables $(M_\text{b},Q_\text{b},\ell)$ changed by the absorption. The conserved quantities of the particle infinitesimally change variables into $(M_\text{b}+dM_\text{b},Q_\text{b}+dQ_\text{b},\ell+d\ell)$. Then, owing to these changes, the locations of minimum point and outer horizon both are infinitesimally moved to $r_\text{min}\rightarrow r_\text{min}+dr_\text{min}$ and $r_h\rightarrow r_h+dr_h$. Under these changes, the configuration of the black hole after the absorption can be expected from a change of the minimum value of the function $df_\text{min}$. Then, the moved minimum point satisfies
\begin{align}
\partial_{r} f(M_\text{b}+dM_\text{b},Q_\text{b}+dQ_b,\ell+d\ell,r)|_{r=r_\text{min}+dr_\text{min}}=f'_\text{min}+df'_\text{min}=0,
\end{align}
which is in terms of variables with partial derivatives
\begin{align}
df'_\text{min}=\frac{\partial f'_\text{min}}{\partial M_\text{b}}dM_\text{b}+\frac{\partial f'_\text{min}}{\partial Q_\text{b}}dQ_\text{b}+\frac{\partial f'_\text{min}}{\partial \ell}d\ell+\frac{\partial f'_\text{min}}{\partial r_\text{min}}dr_\text{min}=0,
\end{align}
with
\begin{align}
\frac{\partial f'_\text{min}}{\partial M_\text{b}}=\frac{16(D-3)\pi}{(D-2)\Omega_{D-2}r_\text{min}^{D-2}},\quad \frac{\partial f'_\text{min}}{\partial Q_\text{b}}=-\frac{16(2D-6)\pi Q}{(D-2)\Omega_{D-2}r_\text{min}^{2D-5}},\quad \frac{\partial f'_\text{min}}{\partial \ell}=-\frac{4 r_\text{min}}{\ell^3}.
\end{align}
The value of the minimum at $r_\text{min}+dr_\text{min}$ becomes
\begin{align}
f(M_b+dM_b,Q_b+dQ_b,\ell+d\ell,r)|_{r=r_\text{min}+dr_\text{min}}&=f_\text{min}+df_\text{min}\\
&=\delta+\left(\frac{\partial f_\text{min}}{\partial M_b}dM_b+\frac{\partial f_\text{min}}{\partial Q_b}dQ_b+\frac{\partial f_\text{min}}{\partial \ell}d\ell\right),\nonumber
\end{align}
where we use $f'_\text{min}=0$ in Eq.\,(\ref{eq:cosmicExtremal}). Then, we can obtain $(dM_\text{b}, d\ell)$ in terms of the particle charges $(q,|p^r|)$ under the particle absorption. Owing to the location of the absorption, the outer horizon, the value of the minimum is obtained under $(r_\text{min},r_\text{h})$. However, this is too complex to analyze and write. Instead, we can impose the condition for the near-extremal black hole that
\begin{align}
\delta \rightarrow \delta_\epsilon,\quad r_\text{h}\rightarrow r_\text{min}+\epsilon.
\end{align}
The outer horizon of the near-extremal black hole is located slightly to the right of the minimum point, and the minimum value is a very small negative value. This is given as $|\delta_\epsilon|,\,\epsilon \ll 1$. For the near-extremal black hole, the moved minimum value is
\begin{align}
f_\text{min}+df_\text{min}=&\left(\delta_\epsilon+\frac{32\pi r_\text{min}^5 (-1-(D-2)r_\text{min}^{1-2D}(-Q^2 r_\text{min}^3+M r_\text{min}^D)\ell^2)|p^r|}{\Omega_{D-2}(D-3)(D-2)(r_\text{min}^{D+2}-2M r_\text{min}^3 \ell^2 + 2 r_\text{min}^D \ell^2)}\right)+\mathcal{O}(\epsilon),
\end{align} 
where we skip to write the first order of $\epsilon$. To simplify this expansion, if we remove $Q^2$ term by using $f'_\text{min}=0$, then, this becomes
\begin{align}
f_\text{min}+df_\text{min}=\delta_\epsilon + \mathcal{O}(\epsilon^2).
\end{align}
The first order of $\epsilon$ is also removed, and the minimum value of the extremal black hole is
\begin{align}
f_\text{min}+df_\text{min}=0,\quad \delta_\epsilon=0,\quad \epsilon=0.
\end{align}
Therefore, extremal and near-extremal black holes stay at their minimum, so their phases cannot be changed, even if they are charged or discharged by the charged particle absorption. This result is quite different from that in cases of no $PV_\text{b}$ term where the extremal black holes are easily broken into a non-extremal one by the absorption. Because the extremal black hole does not change its minimum value, the varied extremal black hole can stay on the $\delta=0$ surface as shown in Fig.\,\ref{fig:f2b}\,(a).
Further, from Eq.\,(\ref{eq:extremaldS01}), to decrease the entropy, the extremal black hole should be contracted under the particle absorption in Fig.\,\ref{fig:f2b}\,(b). On the plane of $(Q,M,\ell)$, the phase of the extremal black hole moves in a three-dimensional direction, because the AdS radius is changed by the absorption. However, as shown in the $M-Q$ diagram of Fig.\,\ref{fig:f2b}\,(b), the extremal black hole stays on the extremal line. The near-extremal black hole also moves its own value of $\delta$.

\section{Summary}\label{sec5}

We investigated variations of the charged AdS black hole under the charged particle absorption by considering the pressure as a cosmological constant. It is known that the thermal conjugate of the pressure is the volume of the black hole inside of the horizon. However, the cosmological constant is not a variable in the action and equations of the motion, so the dynamical effect is not easy to predict with the pressure term. To elucidate the effect of the pressure and volume terms, we consider an infinitesimal variation of the black hole by a charged particle. This is almost the only way to demonstrate the effect of an external particle without using entire equations of the motion. Then, when the particle is absorbed into the black hole, the black hole varies by as much as the conserved quantities of the charged particle. The change of the black hole exactly corresponds to the first law of thermodynamics in terms of the enthalpy. However, the second law of thermodynamics is violated for the extremal and near-extremal black holes in which the entropy decreases under the absorption. It implies that, at least, the entropy of the black hole needs a correction which should not be proportional to the outer horizon. The violation of the second law of thermodynamics can be related to the weak cosmic censorship conjecture which is related to the stability of the horizon. The stability can be shown from the change of the minimum value of the function $f(M,Q,\ell,r)$ under the absorption. Interestingly, the variation of the minimum value is quite different from that in the case without the pressure term. The minimum value of the function $f(M,Q,\ell,r)$ is not changed for extremal and near-extremal black holes under the absorption. It implies that the extremal or near-extremal black holes still stay as they are after an absorption of the external particle. Thus, even if extremal or near-extremal black holes are charged or discharged by the absorption, they maintain their extremality or near-extremality. This result cannot be seen in the charged particle absorption without the thermodynamic pressure and volume term. In addition, owing to the maintenance of its minimum value, the extremal black hole cannot be overcharged in the process. This ensures the stability of the horizon under the charged particle absorption.

\vspace{10pt} 
\newpage
{\bf Acknowledgments}

This work was supported by Basic Science Research Program through the National Research Foundation of Korea (NRF) funded by the Ministry of Science, ICT \& Future Planning (NRF-2015R1C1A1A02037523).

\end{document}